\documentstyle[pra,aps,psfig]{revtex}
\def\Im{\mathop{\rm Im}} 
\def\Tr{\mathop{\rm Tr}} 

\begin{document}
\title{Motional Casimir force}
\author{Marc Thierry Jaekel $^{(a)}$ and Serge Reynaud $^{(b)}$}
\address{(a) Laboratoire de Physique Th\'{e}orique de l'ENS
\thanks{%
Unit\'e propre du Centre National de la Recherche Scientifique,
associ\'ee \`a l'Ecole Normale Sup\'erieure et \`a l'Universit\'e
Paris-Sud}, 24 rue Lhomond, F75231 Paris Cedex 05 France\\
(b) Laboratoire de Spectroscopie Hertzienne
\thanks{%
Unit\'e de l'Ecole Normale Sup\'erieure et de l'Universit\'e
Pierre et Marie Curie, associ\'ee au Centre National de la Recherche 
Scientifique}, 4 place Jussieu, case 74, F75252 Paris Cedex 05 France}
\date{{\sc Journal de Physique I} {\bf 2} (1992) 149-165}
\maketitle

\begin{abstract}
We study the situation where two point like mirrors are placed in the vacuum
state of a scalar field in a two-dimensional spacetime. Describing the
scattering upon the mirrors by transmittivity and reflectivity functions
obeying unitarity, causality and high frequency transparency conditions, we
compute the fluctuations of the Casimir forces exerted upon the two
motionless mirrors. We use the linear response theory to derive the motional
forces exerted upon one mirror when it moves or when the other one moves. We
show that these forces may be resonantly enhanced at the frequencies
corresponding to the cavity modes. We interpret them as the mechanical
consequence of generation of squeezed fields.

\end{abstract}

\section*{Introduction}

The vacuum fluctuations of the electromagnetic field manifest themselves
through macroscopic Casimir forces \cite{Motional1,Motional2,Motional3}.
These forces result from the radiation pressure exerted upon boundaries by
the scattered fluctuations which depends upon the instantaneous and local
value of the stress tensor. Consequently, they are fluctuating quantities 
\cite{Motional4}. More precisely, the forces exerted upon the two mirrors of
a Fabry-Perot cavity have to be considered as random variables. In the
present paper, we compute the associated correlation functions. For the sake
of simplicity, we study the situation where two point like mirrors are
placed in the vacuum state of a scalar field in a two-dimensional (2D)
spacetime.

As illustrated by the Langevin theory of Brownian motion \cite{Motional5},
any fluctuating force has a long term cumulative effect. This cumulative
force can be derived from linear response theory \cite{Motional6}. The
fluctuations of Casimir forces thus imply that mirrors moving in the vacuum
must experience systematic forces. In a previous paper \cite{Motional7}, we
have computed such a motional force for a single mirror in the vacuum state
(of a scalar field in a 2D spacetime). At the limit of perfect reflectivity,
we found a dissipative force proportional to the third time derivative of
the mirror's position $q$ 
\begin{equation}
F\left( t\right) =\frac{\hbar q^{\prime \prime \prime }\left( t\right) }
{6\pi c^2 }  \eqnum{1a}
\end{equation}
which corresponds to a linear susceptibility at the frequency $\omega $ 
\begin{equation}
\chi [\omega ]=\frac{i\hbar \omega ^{3}}{6\pi c^2 }  \eqnum{1b}
\end{equation}
(from now on, we use natural units where $c=1$; however, we keep $\hbar $ as
a scale for vacuum fluctuations). This damping force for a single moving
mirror is connected to the Casimir force (mean force between two mirrors at
rest) since both result from a modification of the vacuum stress tensor \cite
{Motional8}. Actually, expression (1a) identifies with the linear
approximation of the force computed using the techniques of quantum field
theory \cite{Motional9,Motional10,Motional11}. As required by Lorentz
invariance of the vacuum state \cite{Motional12}, the damping force vanishes
for a motion with a uniform velocity. In the case of a uniform acceleration,
the mirror is submitted to the same fluctuating field as if it were at rest
in a thermal field \cite{Motional13} so that it experiences also a zero
motional force.

In the present paper, we compute the explicit expressions of the forces
exerted upon one mirror due either to its own motion (in presence of the
second mirror) or to the motion of the other. These expressions, obtained
from linear response theory, are valid in a first order expansion in the
mirrors' displacement without restriction on the motion's frequency.

A problem in any calculation of vacuum induced effects is to dispose of the
divergences associated with the infiniteness of the total vacuum energy.
This problem can be solved by assuming that the boundaries are transparent
at high frequencies. Using a scattering approach where the mirrors are
described by transmittivity and reflectivity functions obeying unitarity,
causality and high frequency transparency conditions, one obtains a regular
expression for the mean force between two mirrors \cite{Motional3}. The same
approach also provides directly a causal motional force in the single mirror
problem whereas the non causal expression (1) is recovered as an asymptotic
limit for a perfectly reflecting mirror \cite{Motional7}.

In the present paper, we use this approach to study the motional forces in
the two mirrors problem. First, we compute the correlation functions
characterizing the fluctuating Casimir forces exerted upon the two
motionless mirrors. Then, we use the linear response theory to derive the
susceptibility functions associated with the motional forces. In order to
obtain these functions, we use some analytic properties of the correlation
functions which are analysed in Appendix A. We check that our results are
consistent with already known limiting cases: the static Casimir force \cite
{Motional3} (limit of a null frequency), the one mirror problem \cite
{Motional7} (limit where one mirror is transparent) and the limit of perfect
reflection \cite{Motional10}.

The expressions obtained for perfectly reflecting mirrors correspond to a
damping force analogous to equation (1) with two differences. First, the
response is delayed because of the time of flight from one mirror to the
other. Second, the motional modification of the vacuum fields is reflected
back by the mirrors. The resulting interference between the different
numbers of cavity roundtrips gives rise to a divergence of the
susceptibility functions. For partially transmitting mirrors, these
functions are regular and describe a resonant enhancement of the motional
Casimir force, which may be large when the Fabry-Perot cavity has a high
finesse \cite{Motional14}. A resonance approximation is used to evaluate the
susceptibility functions in this case.

It is known that mirrors moving with a non uniform velocity squeeze the
vacuum fields \cite{Motional15,Motional16}. This requires that energy and
impulsion be exchanged between the mirrors and the fields. The motional
Casimir forces thus appear as a mechanical consequence of this squeezing
effect. In Appendix B, we discuss the motional modifications of the field
scattering by the mirrors and we write an effective Hamiltonian describing
the squeezing effect as well as the mechanical forces upon the mirrors.

\section*{Notations}

Any function $f(t)$ defined in the time domain and its Fourier transform 
$f[\omega ]$ are supposed to be related through \footnote{{\it The notation
used in the original paper for Fourier transforms has been changed to 
a more convenient one.}} 
\begin{equation}
f(t)=\int \frac{{\rm d}\omega }{2\pi }f[\omega ]e^{-i\omega t}  \eqnum{2a}
\end{equation}
In the following, some functions of time will be expressed as integrals over
two frequencies, with the notation 
\begin{equation}
f(t)=\int \frac{{\rm d}\omega }{2\pi }\int \frac{{\rm d}\omega ^\prime }
{2\pi }e^{-i\omega t-i\omega ^\prime t}f[\omega ,\omega ^\prime ] 
\eqnum{2b}
\end{equation}
Comparing with (2a), one gets the equivalent expression 
\begin{equation}
f[\omega ]=\int \frac{{\rm d}\omega ^\prime }{2\pi }f[\omega -\omega
^\prime ,\omega ^\prime ]  \eqnum{2c}
\end{equation}

In a 2D spacetime (time coordinate $t$, space coordinate $x$), a free scalar
field is the sum of two counterpropagating components $\varphi (t-x)+\psi
(t+x)$ which will be written as the two components of a column matrix 
\[
\Phi _{x}(t)=\left( 
\begin{array}{c}
\varphi (t-x) \\ 
\psi (t+x)
\end{array}
\right) 
\]

The Fourier transform $\Phi _{x}[\omega ]$ of the column $\Phi _{x}(t)$ is
related to the standard annihilation and creation operators corresponding to
the two propagation directions 
\begin{eqnarray*}
&&\Phi _{x}[\omega ]=\left( 
\begin{array}{c}
\varphi [\omega ]e^{i\omega x} \\ 
\psi [\omega ]e^{-i\omega x}
\end{array}
\right) =e^{i\eta \omega x}\Phi [\omega ] \\
&&\varphi [\omega ]=\sqrt{\frac{\hbar }{2|\omega |}}\left( \theta (\omega
)a_\omega +\theta (-\omega )a_{-\omega }^\dagger \right) \qquad \psi
[\omega ]=\sqrt{\frac{\hbar }{2|\omega |}}\left( \theta (\omega )b_{\omega
}+\theta (-\omega )b_{-\omega }^\dagger \right)
\end{eqnarray*}
where the abbreviated notation $\Phi $ stands for the values of $\Phi _{x}$
evaluated at $x=0$ and 
\[
\eta =\left( 
\begin{array}{cc}
1 & 0 \\ 
0 & -1
\end{array}
\right) 
\]

The energy and impulsion densities correspond to two counterpropagating
energy fluxes 
\[
e_{x}(t)=\varphi ^\prime (t-x)^2 +\psi ^\prime (t+x)^2 \qquad
p_{x}(t)=\varphi ^\prime (t-x)^2 -\psi ^\prime (t+x)^2  
\]
They may be written as integrals over two frequencies (see eqs 2) with 
\[
e_{x}[\omega ,\omega ^\prime ]=i\omega i\omega ^\prime \ \Tr \left[
\Phi _{x}[\omega ]\Phi _{x}[\omega ^\prime ]^{\rm T}\right] \qquad
e_{x}[\omega ,\omega ^\prime ]=i\omega i\omega ^\prime \ \Tr \left[
\eta \Phi _{x}[\omega ]\Phi _{x}[\omega ^\prime ]^{\rm T}\right] 
\]
$\Tr $ stands for the trace operation and $X^{\rm T}$ for the
transposed of $X$. So, their mean values in a quantum state are functions of
the covariance matrix 
\[
\left\langle \Phi _{x}[\omega ]\Phi _{x^\prime }[\omega ^\prime ]^{\rm T}
\right\rangle =e^{i\eta \omega x}\left\langle \Phi [\omega ]\Phi [\omega
^\prime ]^{\rm T}\right\rangle e^{i\eta \omega ^\prime x^\prime } 
\]

For a stationary state such as the vacuum, the covariance matrix depends
only upon one frequency parameter 
\begin{equation}
\left\langle \Phi [\omega ]\Phi [\omega ^\prime ]^{\rm T}\right\rangle
=2\pi \delta (\omega +\omega ^\prime )c[\omega ]  \eqnum{3a}
\end{equation}
It will be useful to write it in terms of the anticommutators which
characterize the field states and of the commutators which do not depend
upon the states 
\begin{eqnarray}
&&c[\omega ]=c_{+}[\omega ]+c_{-}[\omega ]  \eqnum{3b} \\
c_{+}[\omega ] &=&\frac{c[\omega ]+c[-\omega ]^{\rm T}}2 =c_{+}[-\omega
]^{\rm T}\qquad c_{-}[\omega ]=\frac{c[\omega ]-c[-\omega ]^{\rm T}}2 =
\frac{I\hbar }{4\omega }  \nonumber
\end{eqnarray}
$I$ is the unit matrix. The vacuum state corresponds to 
\begin{equation}
c[\omega ]=I\theta (\omega )\frac{\hbar }{2\omega }\qquad c_{+}[\omega
]=I\varepsilon (\omega )\frac{\hbar }{4\omega }\qquad \varepsilon (\omega
)=\theta (\omega )-\theta (-\omega )  \eqnum{4}
\end{equation}

\section*{Scattering upon motionless mirrors}

The scattering of the field upon a partially transmitting mirror at rest at 
$q$ is described by 
\begin{equation}
\Phi _{\rm out}[\omega ]=e^{-i\eta \omega q}S[\omega ]e^{i\eta \omega
q}\Phi _{\rm in}[\omega ]\qquad S[\omega ]=\left( 
\begin{array}{cc}
s[\omega ] & r[\omega ] \\ 
r[\omega ] & s[\omega ]
\end{array}
\right)  \eqnum{5}
\end{equation}
The $S-$matrix is supposed to be real in the temporal domain, causal and
unitary 
\begin{eqnarray}
&&S[-\omega ]=S[\omega ]^{*}  \eqnum{6a} \\
&&S[\omega ]{\rm \ is\ analytic\ (and\ regular)\ for\ }
\Im \omega >0  \eqnum{6b} \\
&&S[\omega ]S[\omega ]^\dagger =1  \eqnum{6c}
\end{eqnarray}
Finally, the mirror is supposed to be transparent at high frequencies 
\begin{equation}
S[\omega ]\rightarrow I{\rm \quad for\ }\omega \rightarrow \infty  \eqnum{6d}
\end{equation}
This assumption will allow us to regularize the ultraviolet divergences
associated with the infiniteness of the vacuum energy \cite{Motional3}. A
mirror perfectly reflecting ($s=0$ and $r=-1$) at all frequencies does not
obey this condition and it will be better to consider the perfect mirror as
the limit of a model obeying the transparency condition (for example a
mirror perfectly reflecting at frequencies below a reflection cutoff).

We study now the situation where two mirrors are placed at rest in the
vacuum at positions $q_1 $ and $q_2 $ (see Figure 1). 
\begin{figure}[h]
\centerline{\psfig{figure=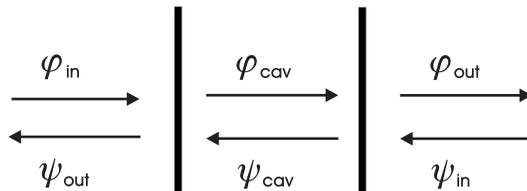,width=7cm}}
\caption{Two mirrors scatter the two counterpropagating fields. The
subscripts `in', `out' and `cav' refer respectively to the input, output and
intracavity fields.}
\label{Fig1}
\end{figure}
As the final results will depend only on the distance $q=q_2 -q_1 $
between the two mirrors, we consider from now on 
\begin{equation}
q_2 =\frac{q}2 \qquad q_1 =-\frac{q}2   \eqnum{7}
\end{equation}

The scattering of the field by the mirrors (see Figure 1) is described by
two scattering matrices given by equations (5) 
\begin{eqnarray*}
\left( 
\begin{array}{c}
\varphi _{\rm cav}[\omega ] \\ 
\psi _{\rm out}[\omega ]
\end{array}
\right) &=&\left( 
\begin{array}{cc}
s_1 [\omega ] & r_1 [\omega ]e^{i\omega q} \\ 
r_1 [\omega ]e^{-i\omega q} & s_1 [\omega ]
\end{array}
\right) \left( 
\begin{array}{c}
\varphi _{\rm in}[\omega ] \\ 
\psi _{\rm cav}[\omega ]
\end{array}
\right) \\
\left( 
\begin{array}{c}
\varphi _{\rm out}[\omega ] \\ 
\psi _{\rm cav}[\omega ]
\end{array}
\right) &=&\left( 
\begin{array}{cc}
s_2 [\omega ] & r_2 [\omega ]e^{-i\omega q} \\ 
r_2 [\omega ]e^{i\omega q} & s_2 [\omega ]
\end{array}
\right) \left( 
\begin{array}{c}
\varphi _{\rm cav}[\omega ] \\ 
\psi _{\rm in}[\omega ]
\end{array}
\right)
\end{eqnarray*}
As usual in the computation of a Fabry Perot cavity, one can solve these
equations to express the outcoming and the intracavity fields in terms of
the input ones 
\[
\Phi _{\rm out}[\omega ]=S[\omega ]\Phi _{\rm in}[\omega ]\qquad \Phi _
{\rm cav}[\omega ]=R[\omega ]\Phi _{\rm in}[\omega ] 
\]
The global scattering matrix $S[\omega ]$ and the resonance matrix $R[\omega
]$ are given by (all the reflectivities and transmittivities are evaluated
at frequency $\omega $) 
\begin{eqnarray}
&&S[\omega ]=\frac1 {d[\omega ]}\left( 
\begin{array}{cc}
s_1 s_2  & r_2 e^{-i\omega q}+d_2 r_1 e^{i\omega q} \\ 
r_1 e^{-i\omega q}+d_1 r_2 e^{i\omega q} & s_1 s_2 
\end{array}
\right)  \eqnum{8} \\
&&R[\omega ]=\frac1 {d[\omega ]}\left( 
\begin{array}{cc}
s_1  & s_2 r_1 e^{i\omega q} \\ 
s_1 r_2 e^{i\omega q} & s_2 
\end{array}
\right)  \eqnum{9} \\
&&d_1 [\omega ]=s_1 [\omega ]^2 -r_1 [\omega ]^2 \qquad d_2 [\omega
]=s_2 [\omega ]^2 -r_2 [\omega ]^2   \nonumber \\
&&d[\omega ]=1-r[\omega ]e^{2i\omega q}\qquad r[\omega ]=r_1 [\omega
]r_2 [\omega ]  \nonumber
\end{eqnarray}
One checks from conditions (6) that the matrices $S$ and $R$ are real in the
time domain, that they are retarded response functions, that the matrix $S$
is unitary and that the Fabry-Perot is transparent at the high frequency
limit 
\begin{eqnarray}
&&S[-\omega ]=S[\omega ]^{*}\qquad R[-\omega ]=R[\omega ]^{*}  \eqnum{10a} \\
&&S[\omega ]{\rm \ and\ }R[\omega ]{\rm \ are\ analytic\ (and\ regular)\
for\ }
\Im \omega >0  \eqnum{10b} \\
&&S[\omega ]S[\omega ]^\dagger =1  \eqnum{10c} \\
&&S[\omega ]\rightarrow I{\rm \quad }R[\omega ]\rightarrow {\rm I\quad for\ }
\omega \rightarrow \infty  \eqnum{10d}
\end{eqnarray}

\section*{Casimir force between two motionless mirrors}

As usual in a local formulation of the Casimir effect \cite{Motional17}, the
forces $F_1 (t)$ and $F_2 (t)$ acting upon the two mirrors can be deduced
from the stress tensor evaluated at both sides of the mirrors 
\begin{eqnarray*}
F_1 (t) &=&\varphi _{\rm in}^\prime (t-q_1 )^2 -\varphi _{\rm cav}
^\prime (t-q_1 )^2 +\psi _{\rm out}^\prime (t+q_1 )^2 -\psi _
{\rm cav}^\prime (t+q_1 )^2  \\
F_2 (t) &=&\varphi _{\rm cav}^\prime (t-q_2 )^2 -\varphi _{\rm out}
^\prime (t-q_2 )^2 +\psi _{\rm cav}^\prime (t+q_2 )^2 -\psi _
{\rm in}^\prime (t+q_2 )^2 
\end{eqnarray*}
The forces are quadratic forms of the vacuum fields as the energy or
impulsion densities, so that they can be expressed as integrals (2) over two
frequencies with (using equation 7) 
\begin{eqnarray*}
F_1 [\omega ,\omega ^\prime ] &=&\exp \left( -\frac{i}2 (\omega +\omega
^\prime )q\right) i\omega i\omega ^\prime \left( \varphi _{\rm in}
[\omega ]\varphi _{\rm in}[\omega ^\prime ]-\varphi _{\rm cav}[\omega
]\varphi _{\rm cav}[\omega ^\prime ]\right) \\
&&+\exp \left( \frac{i}2 (\omega +\omega ^\prime )q\right) i\omega
i\omega ^\prime \left( \psi _{\rm out}[\omega ]\psi _{\rm out}[\omega
^\prime ]-\psi _{\rm cav}[\omega ]\psi _{\rm cav}[\omega ^{\prime
}]\right) \\
F_2 [\omega ,\omega ^\prime ] &=&\exp \left( -\frac{i}2 (\omega +\omega
^\prime )q\right) i\omega i\omega ^\prime \left( \psi _{\rm cav}
[\omega ]\psi _{\rm cav}[\omega ^\prime ]-\psi _{\rm in}[\omega ]\psi
_{\rm in}[\omega ^\prime ]\right) \\
&&+\exp \left( \frac{i}2 (\omega +\omega ^\prime )q\right) i\omega
i\omega ^\prime \left( \varphi _{\rm cav}[\omega ]\varphi _{\rm cav}
[\omega ^\prime ]-\varphi _{\rm out}[\omega ]\varphi _{\rm out}
[\omega ^\prime ]\right)
\end{eqnarray*}
Noting that the matrices 
\[
P_{\pm }=\frac{1\pm \eta }2  
\]
are the projectors onto the counterpropagating components $\varphi $ and 
$\psi $ and using the expressions of the output and cavity fields in terms of
the input ones, one obtains 
\begin{equation}
F_{i}[\omega ,\omega ^\prime ]=i\omega i\omega ^\prime \ \Tr \left[ 
{\cal F}_{i}[\omega ,\omega ^\prime ]\Phi _{x}[\omega ]\Phi _{x}[\omega
^\prime ]^{\rm T}\right]  \eqnum{11a}
\end{equation}
with 
\begin{eqnarray}
{\cal F}_{i}[\omega ,\omega ^\prime ] &=&\varepsilon _{i}\exp \left( -
\frac{i}2 (\omega +\omega ^\prime )q\right) \left( P_{\varepsilon
_{i}}-R[\omega ^\prime ]^{\rm T}P_{\varepsilon _{i}}R[\omega ]\right) 
\eqnum{11b} \\
&&+\varepsilon _{i}\exp \left( \frac{i}2 (\omega +\omega ^{\prime
})q\right) \left( S[\omega ^\prime ]^{\rm T}P_{-\varepsilon
_{i}}S[\omega ]-R[\omega ^\prime ]^{\rm T}P_{-\varepsilon _{i}}R[\omega
]\right)  \nonumber \\
\varepsilon _1  &=&1\qquad \varepsilon _2 =-1  \nonumber
\end{eqnarray}
The two matrices ${\cal F}_{i}$ obey the following properties 
\begin{equation}
{\cal F}_{i}[\omega ,\omega ^\prime ]^{\rm T}={\cal F}_{i}[\omega
^\prime ,\omega ]\qquad {\cal F}[\omega ,\omega ^\prime ]^\dagger =
{\cal F}[-\omega ^\prime ,-\omega ]  \eqnum{12}
\end{equation}

When evaluating the mean forces in the vacuum, one obtains from equations
(3,4) 
\[
\left\langle F_{i}[\omega ,\omega ^\prime ]\right\rangle =\theta (\omega )
\frac{\hbar \omega }2 2\pi \delta (\omega +\omega ^\prime )\ \Tr 
{\cal F}_{i}[\omega ,-\omega ] 
\]
Using the unitarity of the $S-$matrix and the following expression of 
$RR^\dagger $ 
\begin{eqnarray}
&&R[\omega ]R[\omega ]^\dagger =I+Q[\omega ]+Q[\omega ]^\dagger  
\eqnum{13a} \\
&&Q[\omega ]=\frac1 {d[\omega ]}\left( 
\begin{array}{cc}
r_1 r_2 e^{2i\omega q} & r_1 e^{i\omega q} \\ 
r_2 e^{i\omega q} & r_1 r_2 e^{2i\omega q}
\end{array}
\right)  \eqnum{13b}
\end{eqnarray}
one shows that 
\[
\Tr {\cal F}_{i}[\omega ,-\omega ]=\varepsilon _{i}\Tr \left(
I-R[\omega ]R[\omega ]^\dagger \right) =-\varepsilon _{i}\Tr \left(
Q[\omega ]+Q[\omega ]^\dagger \right) 
\]
This leads to the known expression for the Casimir force between two
partially transmitting mirrors \cite{Motional3} 
\begin{equation}
\left\langle F_{i}(t)\right\rangle =-\varepsilon _{i}\int_0 ^{\infty }\frac{
{\rm d}\omega }{2\pi }\hbar \omega \left( \frac{r[\omega ]e^{2i\omega q}}{
d[\omega ]}+\frac{r[\omega ]^{*}e^{-2i\omega q}}{d[\omega ]^{*}}\right) 
\eqnum{14}
\end{equation}

\section*{Force fluctuations}

We now compute the correlation functions characterizing the fluctuations of
the forces $F_1 $ and $F_2 $ acting upon the two mirrors 
\[
C_{ij}(t)=\left\langle F_{i}(t)F_{j}(0)\right\rangle -\left\langle
F_{i}\right\rangle \left\langle F_{j}\right\rangle 
\]
Inserting the expressions (11) of the forces in terms of the field
operators, it appears that the correlation functions $C_{ij}$ are related to
fourth order moments or second order moments of annihilation or creation
operators. With the help of Wick's rules, the fourth order moments may be
deduced from the second order ones. In simple words, the vacuum fields may
be considered as stationary gaussian random variables and higher order
statistical quantities can be deduced from the covariance matrix.

Using this method, simple algebraic manipulations lead to an expression of
the functions $C_{ij}$ as integrals over two frequencies (see equations 2)
with 
\begin{equation}
C_{ij}[\omega ,\omega ^\prime ]=2\omega ^2 \omega ^{\prime \ 2}\ \Tr 
\left[ {\cal F}_{i}[\omega ,\omega ^\prime ]c_{\rm in}[\omega ]{\cal F}
_{j}[\omega ,\omega ^\prime ]^\dagger c_{\rm in}[\omega ^\prime ]^
{\rm T}\right]  \eqnum{15}
\end{equation}
This becomes for the vacuum state (see equation 4) 
\begin{eqnarray}
C_{ij}[\omega ,\omega ^\prime ] &=&\frac{\hbar ^2 }2 \theta (\omega
)\theta (\omega ^\prime )\omega \omega ^\prime \gamma 
_{ij}[\omega ,\omega ^\prime ]  \nonumber \\
\gamma _{ij}[\omega ,\omega ^\prime ] &=&\Tr \left[ {\cal F}
_{i}[\omega ,\omega ^\prime ]{\cal F}_{j}[\omega ,\omega ^{\prime
}]^\dagger \right]  \eqnum{16}
\end{eqnarray}
In other words, the noise spectra associated with the fluctuations of the
Casimir forces may be written 
\begin{equation}
C_{ij}[\omega ]=\frac{\hbar ^2 }2 \theta (\omega )\int_0 ^\omega \frac{
{\rm d}\omega ^\prime }{2\pi }\omega ^\prime (\omega -\omega ^\prime )
\gamma _{ij}[\omega ^\prime ,\omega -\omega ^\prime ]  \eqnum{17}
\end{equation}
From properties (12), the coefficients $\gamma _{ij}$ obey 
\[
\gamma _{ij}[\omega ,\omega ^\prime ]=\gamma _{ij}[\omega ^{\prime
},\omega ]=\gamma _{ji}[\omega ,\omega ^\prime ]^{*}=\gamma _{ji}[-\omega
,-\omega ^\prime ] 
\]

Equations (16-17) give the correlation functions which characterize the
fluctuations of the Casimir forces exerted upon the two mirrors. The
explicit evaluation of the coefficients $\gamma _{ij}$ in terms of the
scattering coefficients is given in the Appendix A (see equations 27 and 21).

\section*{Motional Casimir forces}

We study now the motional Casimir forces. These forces could be obtained by
analysing the modification of the stress tensor associated with the mirror's
motion \cite{Motional8}. At first order in the mirrors' displacements, they
can also be derived from the fluctuations computed for motionless mirrors by
using the linear response theory \cite{Motional6}. We have used both methods
and checked that they provide the same results, as it is the case in the
single mirror problem \cite{Motional7}. Here, we present the linear response
technique. The main steps of the other method are given in Appendix B.

The classical motion of the mirrors corresponds to an effective perturbation
of the Hamiltonian 
\begin{equation}
\delta H(t)=-\sum_{j}F_{j}(t)\delta q_{j}(t)  \eqnum{18}
\end{equation}
where $F_{j}$ are the force operators exerted upon the two mirrors. The
linear response theory provides the mean motional forces associated with
this perturbation in terms of susceptibility functions $\chi _{ij}$ 
\begin{equation}
\left\langle \delta F_{i}(t)\right\rangle =\sum_{j}{\int }{\rm d}\tau \ \chi
_{ij}(\tau )\delta q_{j}(t-\tau )  \eqnum{19a}
\end{equation}
The susceptibility functions $\chi _{ij}$ may be deduced from the
correlation functions since they are the retarded parts of the mean values
of the force commutators $\xi _{ij}$ 
\begin{eqnarray}
\chi _{ij}(t) &=&2i\theta (t)\xi _{ij}(t)  \eqnum{19b} \\
\xi _{ij}(t) &=&\frac{\left[ F_{i}(t),F_{j}(0)\right] }{2\hbar }=\frac{
C_{ij}(t)-C_{ji}(-t)}{2\hbar }  \eqnum{19c}
\end{eqnarray}
The motional forces will be conveniently characterized by the susceptibility
functions written in the frequency domain (compare with equations 1) 
\begin{equation}
\left\langle \delta F_{i}[\omega ]\right\rangle =\sum_{j}\chi _{ij}[\omega
]\delta q_{j}[\omega ]  \eqnum{19d}
\end{equation}

In a first step, we compute the force commutators from the correlation
functions 
\[
\xi _{ij}[\omega ,\omega ^\prime ]=\frac{C_{ij}[\omega ,\omega ^{\prime
}]-C_{ji}[-\omega ^\prime ,-\omega ]}{2\hbar } 
\]
Using the properties (12) and writing the covariances (15) in terms of
anticommutators and commutators (see equations 3), one shows that 
\[
\xi _{ij}[\omega ,\omega ^\prime ]=\frac{\omega \omega ^\prime }2  \Tr
\left[ {\cal F}_{j}[\omega ,\omega ^\prime ]^\dagger {\cal F}
_{i}[\omega ,\omega ^\prime ]\omega c_{+,{\rm in}}[\omega ]+{\cal F}
_{j}[\omega ^\prime ,\omega ]^\dagger {\cal F}_{i}[\omega ^{\prime
},\omega ]\omega ^\prime c_{+,{\rm in}}[\omega ^\prime ]
\right] 
\]
In the vacuum state, this becomes (see equations 4 and 16) 
\[
\xi _{ij}[\omega ,\omega ^\prime ]=\frac{\hbar \omega \omega ^\prime }{8}
\left( \varepsilon (\omega )+\varepsilon (\omega ^\prime )\right) \gamma
_{ij}[\omega ,\omega ^\prime ] 
\]
Writing the Fourier transform of the force commutator as  
\[
\xi _{ij}[\omega ]=\frac{\hbar }{4}\int_0 ^\omega \frac{{\rm d}\omega
^\prime }{2\pi }\omega ^\prime (\omega -\omega ^\prime ) \gamma 
_{ij}[\omega ^\prime ,\omega -\omega ^\prime ] 
\]
one obtains the simple following relation with the noise spectrum (17) 
\[
C_{ij}[\omega ]=2\hbar \theta (\omega )\xi _{ij}[\omega ] 
\]
The noise spectrum $C_{ij}$ contains only positive frequency components, as
expected for the zero temperature state. This corresponds to the fact that
the vacuum fluctuations can damp the mirrors' motion but cannot excite it 
\cite{Motional18}.

In a second step, we deduce the susceptibility functions as the retarded
part of the force commutators. This derivation relies upon the analytic
properties of the correlation functions and requires a detailed inspection
of the expression of the coefficients $\gamma _{ij}[\omega ,\omega ^{\prime
}]$ in terms of the scattering coefficients. This analysis, presented in
Appendix A, shows that $\xi _{ij}[\omega ]$ is a sum of terms which are
either retarded or advanced functions of $\omega $. The susceptibility
functions $\chi _{ij}[\omega ]$ are obtained by retaining only the retarded
terms (see equations 19). One gets them as integrals over two frequencies
with 
\begin{equation}
\chi _{ij}[\omega ,\omega ^\prime ]=\frac{i\hbar \omega \omega ^\prime }
{4}\left( \varepsilon (\omega )\gamma _{ij}^{R}[\omega ,\omega ^{\prime
}]+\varepsilon (\omega ^\prime )\gamma _{ij}^{R}[\omega ^\prime ,\omega
]\right)  \eqnum{20}
\end{equation}
where the coefficients $\gamma _{ij}^{R}$ are the retarded parts of the
coefficients $\gamma _{ij}$ considered as an analytic function of its
second frequency parameter 
\begin{eqnarray}
\gamma _{11}^{R}[\omega ,\omega ^\prime ] &=&2\frac{\alpha _1 [\omega
,\omega ^\prime ]}{d[\omega ]d[\omega ^\prime ]}+\alpha _1 [\omega
,\omega ^\prime ]\beta _1 [\omega ,\omega ^\prime ]r_2 [\omega ]\frac{
e^{2i\omega q}}{d[\omega ]}r_2 [\omega ^\prime ]\frac{e^{2i\omega
^\prime q}}{d[\omega ^\prime ]}  \nonumber \\
&&+r_1 [\omega ^\prime ]d_1 [\omega ]r_2 [\omega ]\frac{e^{2i\omega q}}
{d[\omega ]}+r_1 [\omega ]d_1 [\omega ^\prime ]r_2 [\omega ^\prime ]
\frac{e^{2i\omega ^\prime q}}{d[\omega ^\prime ]}  \nonumber \\
&&+\left( r_1 [\omega ]^{*}+r_1 [\omega ^\prime ]\right) \left(
r_2 [\omega ]^{*}e^{-2i\omega q}+r_2 [\omega ^\prime ]e^{2i\omega
^\prime q}\right) \frac1 {d[\omega ]d[\omega ^\prime ]}  \nonumber \\
&&+2-\frac1 {d[\omega ]}-\frac1 {d[\omega ^\prime ]}  \eqnum{21a}
\end{eqnarray}

\begin{eqnarray}
\gamma _{21}^{R}[\omega ,\omega ^\prime ] &=&-\alpha _1 [\omega ,\omega
^\prime ]\alpha _2 [\omega ,\omega ^\prime ]\frac{e^{i\omega q}}
{d[\omega ]}\frac{e^{i\omega ^\prime q}}{d[\omega ^\prime ]}-\left(
r_1 [\omega ^\prime ]+r_1 [\omega ]^{*}\right) \left( r_2 [\omega
^\prime ]+r_2 [\omega ]^{*}\right) \frac{e^{-i\omega q}}{d[\omega ]^{*}}
\frac{e^{i\omega ^\prime q}}{d[\omega ^\prime ]}  \eqnum{21b} \\
\alpha _{i}[\omega ,\omega ^\prime ] &=&1-s_{i}[\omega ]s_{i}[\omega
^\prime ]+r_{i}[\omega ]r_{i}[\omega ^\prime ]  \eqnum{21c} \\
\beta _{i}[\omega ,\omega ^\prime ] &=&1-s_{i}[\omega ]s_{i}[\omega
^\prime ]-r_{i}[\omega ]r_{i}[\omega ^\prime ]  \eqnum{21d}
\end{eqnarray}
$\gamma _{22}^{R}$ and $\gamma _{12}^{R}$ are obtained by exchanging the
roles of the two mirrors. It has to be noted that the coefficients $\gamma
_{ij}^{R}$ are not symmetrical in the exchange of $\omega $ and $\omega
^\prime $ so that the function $\chi _{ij}[\omega ]$ is not an integral
restricted to the interval $[0,\omega ]$, in contrast to the functions 
$C_{ij}[\omega ]$ (see equation 17) or $\xi _{ij}[\omega ]$.

Equations (20-21) constitute the main results of this paper. They provide
the expressions of the motional Casimir forces (see equations 19 and 2) for
two partially reflecting mirrors in a linear approximation in the mirrors'
displacements.

\section*{Consistency with known results}

Here, we check that the susceptivity functions (20) are consistent with
already known results. This can be done in three limiting cases.

First, considering that the second mirror is transparent at all frequencies 
($r_2 =0$; $s_2 =1$), one obtains 
\begin{equation}
\chi _{11}[\omega ,\omega ^\prime ]=\frac{i\hbar \omega \omega ^\prime }
2 \left( \varepsilon (\omega )+\varepsilon (\omega ^\prime )\right) \alpha
_1 [\omega ,\omega ^\prime ]  \eqnum{22a}
\end{equation}
and one recovers the known susceptibility for a single partially
transmitting mirror moving in the vacuum \cite{Motional7} 
\begin{equation}
\chi _{11}[\omega ]=i\hbar \int_0 ^\omega \frac{{\rm d}\omega ^\prime }
{2\pi }\omega ^\prime (\omega -\omega ^\prime )\alpha _1 [\omega
^\prime ,\omega -\omega ^\prime ]  \eqnum{22b}
\end{equation}

Then, the quasistatic susceptibilities can be computed from equation (20) 
\begin{eqnarray*}
\chi _{ij}[0] &=&\int \frac{{\rm d}\omega }{2\pi }\chi _{ij}[\omega ,-\omega
] \\
\chi _{ij}[\omega ,-\omega ] &=&\frac{i\hbar \omega ^2 }{4}\varepsilon
(\omega )\left( \gamma _{ij}^{R}[-\omega ,\omega ]-\gamma _{ij}^{R}[\omega
,-\omega ]\right)
\end{eqnarray*}
Using equations (21), one checks that they are consistent with the mean
Casimir force (14) between two motionless mirrors 
\[
\chi _{ij}[0]=2i\hbar \varepsilon _{i}\varepsilon _{j}\int_0 ^{\infty }
\frac{{\rm d}\omega }{2\pi }\omega ^2 \left( \frac{r[\omega ]e^{2i\omega q}}
{d[\omega ]^2 }-\frac{r[\omega ]^{*}e^{-2i\omega q}}{d[\omega ]^{*\ 2}}
\right) =\partial _{q_{j}}\left\langle F_{i}\right\rangle 
\]

Finally, one can consider the limiting case of perfectly reflecting mirrors 
($r_1 =r_2 =-1$; $s_1 =s_2 =0$) where the expressions (21) may be
simplified to 
\begin{eqnarray*}
\gamma _{11}^{R}[\omega ,\omega ^\prime ] &=&2\left( \frac2 {d[\omega
^\prime ]}-1\right) \left( \frac1 {d[\omega ]}+\frac1 {d[-\omega ]}
-1\right) +2 \\
\gamma _{21}^{R}[\omega ,\omega ^\prime ] &=&-4\frac{e^{i\omega ^{\prime
}q}}{d[\omega ^\prime ]}\left( \frac{e^{i\omega q}}{d[\omega ]}+\frac{
e^{-i\omega q}}{d[-\omega ]}\right)
\end{eqnarray*}
Simple calculations then lead to the following expression of the motional
force in the time domain 
\begin{eqnarray}
\left\langle \delta F_1 (t)\right\rangle &=& \frac{\hbar }{6\pi } \left(
\delta q_1 ^{\prime \prime \prime }(t)+\delta q_1 ^{\prime \prime \prime
}(t-2q)+\delta q_1 ^{\prime \prime \prime }(t-4q)+\ldots -\delta
q_2 ^{\prime \prime \prime }(t-q)-\delta q_2 ^{\prime \prime \prime
}(t-3q)-\ldots \right)  \nonumber \\
&&+ \frac{\hbar \pi}{6 q^2} \left( \frac1 2 \delta q_1 ^{\prime
}(t)+\delta q_1 ^\prime (t-2q)+\delta q_1 ^\prime (t-4q)+\ldots
-\delta q_2 ^\prime (t-q)-\delta q_2 ^\prime (t-3q)-\ldots \right)
\eqnum{23}
\end{eqnarray}
It can be checked that this is exactly the linear approximation \cite
{Motional19} of the expression obtained for perfectly reflecting mirrors by
Fulling and Davies \cite{Motional10}. The terms proportional to third time
derivatives appear as generalizing the damping force (1) for a single
perfectly reflecting mirror. Now, the force exerted upon one mirror depends
not only on its own motion but also on the motion of the other one. The
response is delayed due to the time of flight between the two mirrors: the
motional modification of the stress tensor has to propagate from one mirror
to the other in order to exert a force on it. Moreover, the modified stress
tensor is reflected back by the two mirrors. The other terms, proportional
to velocities, are not present in the one mirror problem because of the
Lorentz invariance of the vacuum (see the discussion in the introduction).
They are associated with the existence of a static Casimir force in the two
mirrors problem.

\section*{Resonant enhancement of the motional Casimir force}

The expression (23) leads to a divergence of the susceptibility functions
when evaluated at the frequencies $\frac{k\pi }{q}$ with $k$ integer.
Indeed, the contributions corresponding to different numbers of roundtrips
give rise to a constructive interference at those frequencies. In other
words, the Fabry-Perot can be considered as a mechanical resonator with
resonance frequencies corresponding to the optical resonance frequencies.

For partially transmitting mirrors, the divergences due to perfect
reflection will be regularized. In the last section of this paper, we
evaluate the resonance enhancement at the limiting case of a large quality
factor.

In this case, the reflection delays are much shorter than a roundtrip time
and the reflectivity functions are smoother in the frequency domain than the
Airy function describing the cavity modes. Replacing in (21) the
denominators by geometrical series, we obtain each quantity $\gamma
_{ij}^{R} $ as a sum of components evolving at different frequencies: this
expansion is analogous to the one used in the computation of the static
Casimir force at the large distance approximation \cite{Motional3} 
\begin{eqnarray*}
\gamma _{11}^{R}[\omega ,\omega ^\prime ] &=&2\alpha _1 [\omega ,\omega
^\prime ]\sum_{l,m\geq 0}r[\omega ]^{l}r[\omega ^{\prime
}]^{m}e^{2il\omega q+2im\omega ^\prime q} \\
&&+\alpha _1 [\omega ,\omega ^\prime ]\beta _1 [\omega ,\omega ^{\prime
}]r_2 [\omega ]r_2 [\omega ^\prime ]\sum_{l,m\geq 1}r[\omega
]^{l-1}r[\omega ^\prime ]^{m-1}e^{2il\omega q+2im\omega ^\prime q} \\
&&+\left( r_1 [-\omega ]+r_1 [\omega ^\prime ]\right) r_2 [-\omega
]\sum_{l\geq 1,m\geq 0}r[-\omega ]^{l-1}r[\omega ^{\prime
}]^{m}e^{-2il\omega q+2im\omega ^\prime q} \\
&&+\left( r_1 [-\omega ]+r_1 [\omega ^\prime ]\right) r_2 [\omega
^\prime ]\sum_{l\geq 0,m\geq 1}r[-\omega ]^{l}r[\omega ^{\prime
}]^{m-1}e^{-2il\omega q+2im\omega ^\prime q} \\
&&+r_1 [\omega ^\prime ]d_1 [\omega ]r_2 [\omega ]\sum_{l\geq
1}r[\omega ]^{l-1}e^{2il\omega q} \\
&&+r_1 [\omega ]d_1 [\omega ^\prime ]r_2 [\omega ^\prime ]\sum_{m\geq
1}r[\omega ^\prime ]^{m-1}e^{2im\omega ^\prime q} \\
&&-\sum_{l\geq 1}r[\omega ]^{l}e^{2il\omega q}-\sum_{m\geq 1}r[\omega
^\prime ]^{m}e^{2im\omega ^\prime q}
\end{eqnarray*}
\begin{eqnarray*}
\gamma _{21}^{R}[\omega ,\omega ^\prime ] &=&-\alpha _1 [\omega ,\omega
^\prime ]\alpha _2 [\omega ,\omega ^\prime ]\sum_{l,m\geq 0}r[\omega
]^{l}r[\omega ^\prime ]^{m}e^{i\left( 2l+1\right) \omega q+i\left(
2m+1\right) \omega ^\prime q} \\
&&-\left( r_1 [\omega ^\prime ]+r_1 [-\omega ]\right) \left(
r_2 [\omega ^\prime ]+r_2 [-\omega ]\right) \sum_{l,m\geq 0}r[-\omega
]^{l}r[\omega ^\prime ]^{m}e^{-i\left( 2l+1\right) \omega q+i\left(
2m+1\right) \omega ^\prime q}
\end{eqnarray*}
We will denote $\mu _{ij}^{(L,M)}$ the coefficients in this expansion 
\[
\gamma _{ij}^{R}[\omega ,\omega ^\prime ]=\sum_{L,M}\mu
_{ij}^{(L,M)}[\omega ,\omega ^\prime ]e^{iL\omega q+iM\omega ^\prime q} 
\]
The coefficients $\mu _{ij}^{(L,M)}$ depend only upon the reflectivity
functions. Negative values of $M$ do not appear in the sum because $\mu
_{ij}^{(L,M)}[\omega ,\omega ^\prime ]$ is a retarded function of $\omega
^\prime $. Only even values of $L$ and $M$ appear for $i=j$ and odd ones
for $i\neq j$.

Then, the susceptibility functions $\chi _{ij}[\omega ]$ will be obtained
through an integration (see equations 2) 
\begin{eqnarray}
\chi _{ij}[\omega ] &=&\sum_{L,M}\chi _{ij}^{(L,M)}[\omega ]  \eqnum{24a} \\
\chi _{ij}^{(L,M)}[\omega ] &=&{\int }\frac{{\rm d}\omega ^\prime }{2\pi }
\chi _{ij}^{(L,M)}[\omega ^\prime ,\omega -\omega ^\prime ]  \eqnum{24b}
\\
\chi _{ij}^{(L,M)}[\omega ,\omega ^\prime ] &=&\frac{i\hbar }{4}\omega
\omega ^\prime \left( \varepsilon (\omega )\mu _{ij}^{(L,M)}[\omega
,\omega ^\prime ]e^{iL\omega q+iM\omega ^\prime q}+\varepsilon (\omega
^\prime )\mu _{ij}^{(L,M)}[\omega ^\prime ,\omega ]e^{iL\omega ^{\prime
}q+iM\omega q}\right)  \eqnum{24c}
\end{eqnarray}
It follows that the integrals $\chi _{ij}^{(L,M)}$ with $L\neq M$ will
contain exponentials with a rapidly varying phase and can be considered as
non resonant terms. At the limit of perfect reflection, they provide the
terms proportional to the velocities in equations (23) while the resonant
terms $\chi _{ij}^{(L,M)}$ with $L=M$ provide the terms proportional to the
third time derivatives.

In order to obtain the behaviour of the susceptibility functions near a
resonance at $\frac{k\pi }{q}$ with $k$ a large integer, we will retain only
the terms $L=M$ (the first time derivatives have a small contribution to
equation 23 compared to the third time derivatives in this case). In this
resonance approximation, we obtain 
\begin{eqnarray*}
&&\gamma _{11}^{R}[\omega ,\omega ^\prime ] = \sum_{l\geq 0}\mu
_{11}^{(2l,2l)}[\omega ,\omega ^\prime ]e^{2il(\omega +\omega ^\prime )q}
\\
&&\mu _{11}^{(2l,2l)}[\omega ,\omega ^\prime ] = 2\alpha _1 [\omega
,\omega ^\prime ]r[\omega ]^{l}r[\omega ^\prime ]^{l}+\theta _{l\geq
1}\alpha _1 [\omega ,\omega ^\prime ]\beta _1 [\omega ,\omega ^{\prime
}]r_2 [\omega ]r_2 [\omega ^\prime ]r[\omega ]^{l-1}r[\omega ^{\prime
}]^{l-1} 
\end{eqnarray*}
\begin{eqnarray*}
&&\gamma _{21}^{R}[\omega ,\omega ^\prime ] = \sum_{l\geq 0}\mu
_{21}^{(2l+1,2l+1)}[\omega ,\omega ^\prime ]e^{i(2l+1)(\omega +\omega
^\prime )q} \\
&&\mu _{21}^{(2l+1,2l+1)}[\omega ,\omega ^\prime ] = -\alpha _{1[}\omega
,\omega ^\prime ]\alpha _2 [\omega ,\omega ^\prime ]r[\omega
]^{l}r[\omega ^\prime ]^{l}
\end{eqnarray*}
$\theta _{l\geq 1}=0$ for $l=0$ and $\theta _{l\geq 1}=1$ for $l\geq 1$. As
these expressions are symmetrical in the exchange of the two parameters 
$\omega $ and $\omega ^\prime $ (this was not the case for the general
expressions 21 of $\gamma _{ij}^{R}$), equations (24) lead to 
\begin{eqnarray*}
\chi _{ij}[\omega ] &=&\sum_{L}\chi _{ij}^{(L,L)}[\omega ] \\
\chi _{ij}^{(L,L)}[\omega ] &=&\frac{i\hbar }2 e^{iL\omega q}
{\int_0 ^\omega }\frac{{\rm d}\omega ^\prime }{2\pi }\omega ^{\prime
}(\omega -\omega ^\prime )\mu _{ij}^{(L,L)}[\omega ^\prime ,\omega
-\omega ^\prime ]
\end{eqnarray*}

The various terms $\chi _{ij}^{(L,L)}$ correspond to a motional force
evaluated at a delay time close to a multiple of the time of flight $q$
between the two mirrors. For example, the term $\chi _{ij}^{(0,0)}$
describes the response of the force $F_1 $ to the motion $q_1 $ evaluated
at times much shorter than the roundtrip time. As it could be expected, it
does not depend upon the presence of the second mirror and is the same as if
the mirror 1 were alone (compare $\mu _{11}^{(0,0)}[\omega ,\omega ^{\prime
}]=2\alpha _1 [\omega ,\omega ^\prime ]$ with equation 22). The other
contributions to $\chi _{11}$ correspond to the modification of the damping
force due to the presence of the second mirror. They depend upon the
reflectivities of the two mirrors and appear at time delays close to a
multiple of the roundtrip time $2q$.

The terms $\chi _{21}^{(L,L)}$ describe the force exerted upon one mirror
when the other one moves. They depend upon the reflectivity of both mirrors
and appear at time delays close to an odd multiple of the time of flight $q$.
The lowest order term has a simple form 
\[
\mu _{21}^{(1,1)}[\omega ,\omega ^\prime ]=-\alpha _1 [\omega ,\omega
^\prime ]\alpha _2 [\omega ,\omega ^\prime ] 
\]
The moving mirror modifies the stress tensor of the vacuum field
(modification described by the function $\alpha _1 $). Then the radiation
pressure experienced by the other mirror registers the modification of the
stress tensor (detection efficiency described by the function $\alpha _2 $).

In the limiting case of perfect reflection at all frequencies lower than $%
\omega $, one recovers the terms proportional to the third time derivatives
in equations (23). The susceptibility functions diverge in this case but
they are regular as soon as the mirrors have a small partial transmission.
Indeed, the contributions corresponding to different numbers of roundtrips
can be summed up to give in the resonance approximation 
\begin{eqnarray*}
\gamma _{11}^{R}[\omega ,\omega ^\prime ] &=&\frac{2\alpha _1 [\omega
,\omega ^\prime ]+\alpha _1 [\omega ,\omega ^\prime ]\beta _1 [\omega
,\omega ^\prime ]r_2 [\omega ]r_2 [\omega ^\prime ]e^{2i(\omega
+\omega ^\prime )q}}{D[\omega ,\omega ^\prime ]} \\
\gamma _{21}^{R}[\omega ,\omega ^\prime ] &=&-\frac{\alpha _1 [\omega
,\omega ^\prime ]\alpha _2 [\omega ,\omega ^\prime ]e^{i(\omega +\omega
^\prime )q}}{D[\omega ,\omega ^\prime ]} \\
D[\omega ,\omega ^\prime ] &=&1-r[\omega ]r[\omega ^\prime ]e^{2i(\omega
+\omega ^\prime )q}
\end{eqnarray*}
The denominator $D$ characterizes the resonances of the Fabry Perot cavity
considered as a mechanical resonator.

Considering that the reflectivity coefficients may be approximated as
constant functions from $0$ to $\omega $, the susceptibility functions are
given by the simple expressions ($r=r_1 r_2 $) 
\begin{eqnarray*}
&&\gamma _{11}^{R}[\omega ,\omega ^\prime ]=\frac{4r_1 ^2 }{%
1-r^2 e^{2i(\omega +\omega ^\prime )q}} \\
&&\gamma _{21}^{R}[\omega ,\omega ^\prime ]=-\frac{4r^2 e^{i(\omega
+\omega ^\prime )q}}{1-r^2 e^{2i(\omega +\omega ^\prime )q}} \\
&&\chi _{11}[\omega ]=\frac{i\hbar \omega ^{3}}{6\pi }\frac{r_1 ^2 }
{1-r^2 e^{2i\omega q}} \\
&&\chi _{21}[\omega ]=-\frac{i\hbar \omega ^{3}}{6\pi }\frac{r^2 e^{i\omega
q}}{1-r^2 e^{2i\omega q}}
\end{eqnarray*}
corresponding to a motional force 
\[
\left\langle \delta F_1 (t)\right\rangle =\frac{\hbar }{6\pi }\left(
r_1 ^2 \left( \delta q_1 ^{\prime \prime \prime }(t)+r^2 \delta
q_1 ^{\prime \prime \prime }(t-2q)+r^{4}\delta q_1 ^{\prime \prime \prime
}(t-4q)+\ldots \right) -r^2 \delta q_2 ^{\prime \prime \prime
}(t-q)-r^{4}\delta q_2 ^{\prime \prime \prime }(t-3q)-\ldots \right) 
\]
When compared with equation (23), one notes that the first time derivatives
do not appear here because of the resonance approximation (the expression is
valid only at high enough frequencies $\omega \gg \frac{\pi }{q}$). But the
effect of imperfect reflection of the mirrors is now taken into account 
($r_1 ^2 $ and $r_2 ^2 $ are the reflection coefficients of the two
mirrors for energy densities) and the susceptibilities are regular functions
of the frequency.

\section*{Conclusion}

The motional Casimir force constitutes a new type of interaction between two
mirrors. As the stationary Casimir effect, it is associated with a
modification of the vacuum stress tensor due to the field scattering upon
the mirrors. But it is resonantly enhanced at the resonance frequencies of
the optical cavity $\frac{k\pi }{q}$. When compared with the case of a
single mirror, the enhancement may reach the value $\frac{r_1 ^2 }{%
1-r_1 ^2 r_2 ^2 }$. Consequently, the motional force might be very large 
\cite{Motional20} with the high finesse cavities such as those used in
cavity QED \cite{Motional14}.

\medskip
\noindent {\bf Acknowledgements}

We thank A. Heidmann for discussions.

\appendix

\section{Analytic properties of the correlation functions}

For the sake of clarity, we recall here the expressions of the
susceptibility functions (see equations 2 and 19) 
\begin{eqnarray}
\chi _{ij}(t) &=&2i\theta (t)\xi _{ij}(t)  \eqnum{25a} \\
\xi _{ij}(t) &=&\int \frac{{\rm d}\omega }{2\pi }\int \frac{{\rm d}\omega
^\prime }{2\pi }e^{-i\omega t-i\omega ^\prime t}\xi _{ij}[\omega ,\omega
^\prime ]  \eqnum{25b} \\
\xi _{ij}[\omega ,\omega ^\prime ] &=&\frac{\hbar \omega \omega ^\prime }
{8}\left( \varepsilon (\omega )+\varepsilon (\omega ^\prime )\right)
\gamma _{ij}[\omega ,\omega ^\prime ]  \eqnum{25c}
\end{eqnarray}
Since the susceptibilities are related to the retarded part of the
correlation functions, their derivation relies upon the analytic properties
of the coefficients $\gamma _{ij}$. We show here how these properties can be
inferred from the expressions of the coefficients $\gamma _{ij}$ in terms of
the scattering coefficients, which are themselves analytic functions of the
frequency (see equations 10). The coefficients $\gamma _{ij}$ are obtained
from products of two matrices ${\cal F}_{i}$ (see equation 16) which are
functions of the scattering and resonance matrices $S$ and $R$ (see equation
11b). Developping the corresponding expressions, one obtains rather lengthy
expressions.

However, these expressions may be simplified by using the following
properties. First the $S-$matrix is unitary ($S[\omega ]S[\omega ]^{\dagger
}=I$) and the matrix $R[\omega ]R[\omega ]^\dagger $\ may be written in
terms of the retarded (analytic for $%
\Im \omega >0$) and advanced (analytic for $%
\Im \omega <0$) components $Q[\omega ]$ and $Q[\omega ]^\dagger $\ (see
equations 13). Then, it is also possible to reduce products $R[\omega
]S[\omega ]^\dagger $\ and $S[\omega ]R[\omega ]^\dagger $\ by noting
that they determine the expression of the intracavity fields in terms of the
output ones 
\[
\Phi _{\rm cav}[\omega ]=R[\omega ]S[\omega ]^\dagger \ \Phi _{\rm out}
[\omega ] 
\]
Hence, $R[\omega ]S[\omega ]^\dagger $\ is an advanced response function
and its adjoint $S[\omega ]R[\omega ]^\dagger $\ is a retarded one. Simple
manipulations lead to 
\begin{eqnarray}
S[\omega ]R[\omega ]^\dagger  &=&\overline{R}[\omega ]\qquad R[\omega
]S[\omega ]^\dagger =\overline{R}[\omega ]^\dagger   \eqnum{26a} \\
\overline{R}[\omega ] &=&\frac1 {d[\omega ]}\left( 
\begin{array}{cc}
s_2  & s_2 r_1 e^{i\omega q} \\ 
s_1 r_2 e^{i\omega q} & s_1 
\end{array}
\right)  \eqnum{26b}
\end{eqnarray}

Using these properties, the coefficients $\gamma _{ij}$ are written as sums
of terms, each of them being easily recognized as either a retarded or an 
advanced function of the two frequency parameters $\omega $ and $\omega
^\prime $ 
\begin{eqnarray*}
\gamma _{ij}[\omega ,\omega ^\prime ] &=&\varepsilon _{i}\varepsilon
_{j}\left( \Tr \left[ P_{\varepsilon _{i}}P_{\varepsilon
_{j}}+P_{-\varepsilon _{i}}P_{-\varepsilon _{j}}\right] \right. \\
&&-\Tr \left[ P_{\varepsilon _{i}}R[\omega ]P_{\varepsilon _{j}}R[\omega
^\prime ]^{\rm T}+P_{\varepsilon _{i}}R[\omega ]^{\dagger
}P_{\varepsilon _{j}}R[\omega ^\prime ]^{*}\right] \\
&&-\Tr \left[ P_{-\varepsilon _{i}}\overline{R}[\omega ]P_{-\varepsilon
_{j}}\overline{R}[\omega ^\prime ]^{\rm T}+P_{-\varepsilon _{i}}%
\overline{R}[\omega ]^\dagger P_{-\varepsilon _{j}}\overline{R}[\omega
^\prime ]^{*}\right] \\
&&+\Tr \left[ P_{\varepsilon _{i}}\left( I+Q[\omega ]+Q[\omega
]^\dagger \right) P_{\varepsilon _{j}}\left( I+Q[\omega ^\prime ]^{\rm T}
+Q[\omega ^\prime ]^{*}\right) \right] \\
&&+\Tr \left[ P_{-\varepsilon _{i}}\left( I+Q[\omega ]+Q[\omega
]^\dagger \right) P_{-\varepsilon _{j}}\left( I+Q[\omega ^\prime ]^
{\rm T}+Q[\omega ^\prime ]^{*}\right) \right] \\
&&+e^{i(\omega +\omega ^\prime )q}\Tr \left[ P_{-\varepsilon
_{i}}S[\omega ]P_{\varepsilon _{j}}S[\omega ^\prime ]^{\rm T}\right]
+e^{-i(\omega +\omega ^\prime )q}\Tr \left[ P_{\varepsilon
_{i}}S[\omega ]^\dagger P_{-\varepsilon _{j}}S[\omega ^{\prime
}]^{*}\right] \\
&&-e^{i(\omega +\omega ^\prime )q}\Tr \left[ P_{-\varepsilon
_{i}}R[\omega ]P_{\varepsilon _{j}}R[\omega ^\prime ]^{\rm T}
+P_{-\varepsilon _{i}}\overline{R}[\omega ]P_{\varepsilon _{j}}\overline{R}
[\omega ^\prime ]^{\rm T}\right] \\
&&-e^{-i(\omega +\omega ^\prime )q}\Tr \left[ P_{\varepsilon
_{i}}R[\omega ]^\dagger P_{-\varepsilon _{j}}R[\omega ^{\prime
}]^{*}+P_{\varepsilon _{i}}\overline{R}[\omega ]^\dagger P_{-\varepsilon
_{j}}\overline{R}[\omega ^\prime ]^{*}\right] \\
&&+e^{i(\omega +\omega ^\prime )q}\Tr \left[ P_{-\varepsilon
_{i}}\left( I+Q[\omega ]+Q[\omega ]^\dagger \right) P_{\varepsilon
_{j}}\left( I+Q[\omega ^\prime ]^{\rm T}+Q[\omega ^\prime ]^{*}\right)
\right] \\
&&+\left. e^{-i(\omega +\omega ^\prime )q}\Tr \left[ P_{\varepsilon
_{i}}\left( I+Q[\omega ]+Q[\omega ]^\dagger \right) P_{-\varepsilon
_{j}}\left( I+Q[\omega ^\prime ]^{\rm T}+Q[\omega ^\prime ]^{*}\right)
\right] \right)
\end{eqnarray*}

Now, the susceptibility functions $\chi _{ij}$ are obtained by retaining the
retarded terms and dropping the advanced ones in $\xi _{ij}$ (see equations
25). Consider first the contribution of a term which contains the factor 
$\varepsilon (\omega )$, which is still an analytic function of $\omega
^\prime $ but not of $\omega $. Its integration over $\omega ^\prime $
(see equations 25) provides a contribution to $\chi _{ij}(t)$ which is
either retarded (vanishing for $t<0$) or advanced (vanishing for $t>0$).
Some terms do not depend upon $\omega ^\prime $ and their integration
contributes for a half to retarded terms and for a half to advanced ones as
can be verified explicitly on expressions (25). The terms containing the
factor $\varepsilon (\omega ^\prime )$ are computed in the same manner by
considering the integration over $\omega $.

One finally obtains $\chi _{ij}(t)$ as an integral over two frequencies (see
equations 2) with
\[
\chi _{ij}[\omega ,\omega ^\prime ]=\frac{i\hbar \omega \omega ^\prime }
{4}\left( \varepsilon (\omega )\gamma _{ij}^{R}[\omega ,\omega ^{\prime
}]+\varepsilon (\omega ^\prime )\gamma _{ij}^{R}[\omega ^\prime ,\omega
]\right) 
\]
where $\gamma _{ij}^{R}$ is the retarded part of $\gamma _{ij}$ considered
as a function of its second frequency parameter, that is (assuming $q>0$) 
\begin{eqnarray*}
\gamma _{ij}^{R}[\omega ,\omega ^\prime ] &=&\varepsilon _{i}\varepsilon
_{j}\left( \frac1 2 \Tr \left[ P_{\varepsilon _{i}}P_{\varepsilon
_{j}}+P_{-\varepsilon _{i}}P_{-\varepsilon _{j}}\right] \right. \\
&&-\Tr \left[ P_{\varepsilon _{i}}R[\omega ]P_{\varepsilon _{j}}R[\omega
^\prime ]^{\rm T}\right] -\Tr \left[ P_{-\varepsilon _{i}}\overline{R}
[\omega ]P_{-\varepsilon _{j}}\overline{R}[\omega ^\prime ]^{\rm T}
\right] \\
&&+\Tr \left[ P_{\varepsilon _{i}}\left( I+Q[\omega ]+Q[\omega
]^\dagger \right) P_{\varepsilon _{j}}\left( \frac{I}2 +Q[\omega ^{\prime
}]^{\rm T}\right) \right] \\
&&+\Tr \left[ P_{-\varepsilon _{i}}\left( I+Q[\omega ]+Q[\omega
]^\dagger \right) P_{-\varepsilon _{j}}\left( \frac{I}2 +Q[\omega
^\prime ]^{\rm T}\right) \right] \\
&&+e^{i(\omega +\omega ^\prime )q}\Tr \left[ P_{-\varepsilon
_{i}}S[\omega ]P_{\varepsilon _{j}}S[\omega ^\prime ]^{\rm T}\right] \\
&&-e^{i(\omega +\omega ^\prime )q}\Tr \left[ P_{-\varepsilon
_{i}}R[\omega ]P_{\varepsilon _{j}}R[\omega ^\prime ]^{\rm T}
+P_{-\varepsilon _{i}}\overline{R}[\omega ]P_{\varepsilon _{j}}\overline{R}
[\omega ^\prime ]^{\rm T}\right] \\
&&+e^{i(\omega +\omega ^\prime )q}\Tr \left[ P_{-\varepsilon
_{i}}\left( I+Q[\omega ]+Q[\omega ]^\dagger \right) P_{\varepsilon
_{j}}\left( I+Q[\omega ^\prime ]^{\rm T}\right) \right] \\
&&+\left. e^{-i(\omega +\omega ^\prime )q}\Tr \left[ P_{\varepsilon
_{i}}\left( I+Q[\omega ]+Q[\omega ]^\dagger \right) P_{-\varepsilon
_{j}} Q[\omega ^\prime ]^{\rm T} \right] \right)
\end{eqnarray*}

Using the expressions (8), (9), (13) and (26) of the matrices $S$, $R$, $Q$
and $\overline{R}$ in terms of the scattering coefficients describing the
two mirrors, algebraic manipulations lead to the expressions (21) of the
coefficients $\gamma _{ij}^{R}$.

As discussed previously, the coefficients $\gamma _{ij}$ contain terms
corresponding to retarded contributions and which appear also in the
coefficients $\gamma _{ij}^{R}$. The other terms correspond to advanced
contributions and have been dropped. A comparison between the two types of
terms shows that the coefficients $\gamma _{ij}$ can be deduced in a simple
manner from the coefficients $\gamma _{ij}^{R}$ 
\begin{equation}
\gamma _{ij}[\omega ,\omega ^\prime ]=\gamma _{ij}^{R}[\omega ,\omega
^\prime ]+\gamma _{ji}^{R}[\omega ,\omega ^\prime ]^{*}  \eqnum{27}
\end{equation}
This relation between the correlation function and the susceptibility
function may be considered as the expression of the fluctuation dissipation
theorem \cite{Motional6} for the present problem. With the help of
expressions (21), it provides the explicit expressions of the coefficients 
$\gamma _{ij}$ in terms of the scattering coefficients.

\section{Connection with squeezing}

In the one mirror problem, it has been possible to compute the damping force
by considering that the field scattering is modified when the mirror moves.
It has been shown that this approach is completely equivalent to the linear
response technique \cite{Motional7}. This can also be shown in the two
mirrors problem studied in the present paper.

For a single mirror, the modified scattering matrix $S_{i}$ can be written
in a first order expansion in the mirror's motion $\delta q_{i}$ around the
position $q_{i}$ as \cite{Motional7} 
\begin{eqnarray*}
\Phi _{\rm out}[\omega ] &=&{\int }\frac{{\rm d}\omega ^\prime }{2\pi }
S_{i}[\omega ,\omega ^\prime ]\Phi _{\rm in}[\omega ^{\prime}] \\
S_{i}[\omega ,\omega ^\prime ] &=&2\pi \delta (\omega -\omega
^\prime )\overline{S}_{i}[\omega ]+\delta \overline{S}_{i}[\omega ,\omega
^\prime ] \\
\overline{S}_{i}[\omega ] &=&e^{-i\eta \omega q_{i}}S_{i}[\omega ]e^{i\eta
\omega q_{i}} \\
\delta \overline{S}_{i}[\omega ,\omega ^\prime ] &=&i\omega ^{\prime
}\delta q[\omega -\omega ^\prime ]e^{-i\eta \omega q_{i}}\left(
S_{i}[\omega ]\eta -\eta S_{i}[\omega ^\prime ]\right) e^{i\eta \omega
^\prime q_{i}}
\end{eqnarray*}
The modification $\delta S$ and $\delta R$ of the matrices associated with
the Fabry-Perot can be derived from the elementary matrices $\delta S_{i}$
associated with each mirror. One computes 
\begin{eqnarray*}
\delta R[\omega ,\omega ^\prime ] &=&\frac1 {d[\omega ]}\sum_{i}\left(
P_{\varepsilon _{i}}+P_{-\varepsilon _{i}}\overline{S}_{\overline{i}}[\omega
]P_{\varepsilon _{i}}\right) \delta \overline{S}_{i}[\omega ,\omega ^{\prime
}]\left( P_{\varepsilon _{i}}+P_{-\varepsilon _{i}}R[\omega ^{\prime
}]\right) \\
\delta S[\omega ,\omega ^\prime ] &=&\sum_{i}P_{-\varepsilon _{i}}\delta 
\overline{S}_{i}[\omega ,\omega ^\prime ]\left( P_{\varepsilon
_{i}}+P_{-\varepsilon _{i}}R[\omega ^\prime ]\right)
+\sum_{i}P_{-\varepsilon _{i}}\overline{S}_{i}[\omega ]P_{-\varepsilon
_{i}}\delta R[\omega ,\omega ^\prime ] \\
\overline{i} &=&3-i
\end{eqnarray*}
Then, one obtains the modified mean force through equations analogous to
(11). The susceptibility functions given by the linear response theory are
recovered at the end of lengthy calculations.

This discussion shows that the motional Casimir force is connected to the
problem of squeezing \cite{Motional16}. As a matter of fact, the motional
modification of the field scattering corresponds to a squeezing of the input
fields \cite{Motional15,Motional7}. This squeezing generation requires that
energy and impulsion be exchanged between the field and the mirrors. The
motional Casimir force can be interpreted as a mechanical consequence of
this effect.

As in the single mirror problem, there exists an effective Hamiltonian which
describes the squeezing effect (modification of the field in response to the
mirrors' motion) as well as the Casimir forces (mechanical action upon the
mirrors in response to a variation of the field stress tensor). This
effective Hamiltonian is the secular part (component at zero frequency) of
the coupling (18) 
\[
\delta H[0]={\int }{\rm d}t\ \delta H(t)=\int \frac{{\rm d}\omega }{2\pi }
\int \frac{{\rm d}\omega ^\prime }{2\pi }\sum_{j}\delta q_{j}[-\omega
-\omega ^\prime ]\omega \omega ^\prime \Tr \left[ {\cal F}
_{j}[\omega ,\omega ^\prime ]\Phi _{\rm in}[\omega ]\Phi _{\rm in}
[\omega ^\prime ]^{\rm T}\right] 
\]


\begin{references}
\bibitem{Motional19}  The linear approximation corresponds here to assuming $%
\delta q_{i}\ll q$ and $q_{i}^\prime \ll c=1.$ This suggests that the
domain of validity of the first-order expansion in the mirror's
displacements is defined by these conditions. Fulling and Davies \cite
{Motional10} give only the contribution of the intracavity fields to the
motional force; the contribution of the outer fields, which can also be
derived from their results, has to be added in order to obtain the total
force (23).
\bibitem{Motional20}  In order to estimate the magnitude of this motional
force in realistic experiments, it will be necessary to generalize these
calculations to the situation where two finite area flat mirrors move in the
electromagnetic vacuum in a 4D spacetime.
\end{references}
\end{document}